\documentclass[aps,prl,reprint,superscriptaddress,amssymb,amsmath]{revtex4-1}
\usepackage{graphicx}
\usepackage{dcolumn}
\usepackage{bm, color}
\usepackage[none]{hyphenat} 
\usepackage{bbm}
\usepackage{braket}
\usepackage{tabularx}

\begin{document}

\title{Prediction of a magnetic Weyl semimetal without spin-orbit coupling and strong anomalous Hall effect in the Heusler compensated ferrimagnet Ti$_2$MnAl}
\author{Wujun Shi} 
\thanks{These two authors contributed equally}
\affiliation{Max Planck Institute for Chemical Physics of Solids, D-01187 Dresden, Germany}
\affiliation{School of Physical Science and Technology, ShanghaiTech University, Shanghai 200031, China}
\author{Lukas Muechler} 
\thanks{These two authors contributed equally}
\affiliation{Department of Chemistry, Princeton University, Princeton, New Jersey 08544, USA}
\author{Kaustuv Manna} 
\affiliation{Max Planck Institute for Chemical Physics of Solids, D-01187 Dresden, Germany}
\author{Yang Zhang}
\affiliation{Max Planck Institute for Chemical Physics of Solids, D-01187 Dresden, Germany}
\affiliation{Leibniz Institute for Solid State and Materials Research, 01069 Dresden, Germany}
\author{Klaus Koepernik} 
\affiliation{Leibniz Institute for Solid State and Materials Research, 01069 Dresden, Germany}
\affiliation{IFW Dresden, P.O. Box 270116, D-01171 Dresden, Germany}
\author{Roberto Car}
\affiliation{Department of Chemistry, Princeton University, Princeton, New Jersey 08544, USA}
\author{Jeroen van den Brink}
\affiliation{Leibniz Institute for Solid State and Materials Research, 01069 Dresden, Germany}
\author{Claudia Felser} 
\affiliation{Max Planck Institute for Chemical Physics of Solids, D-01187 Dresden, Germany}
\author{Yan Sun} 
\email{ysun@cpfs.mpg.de}
\affiliation{Max Planck Institute for Chemical Physics of Solids, D-01187 Dresden, Germany}

\date{\today}
\begin{abstract}
We predict a magnetic Weyl semimetal in the inverse Heusler Ti$_2$MnAl, a 
compensated ferrimagnet with a vanishing net magnetic moment and a Curie 
temperature of over 650 K. Despite the vanishing net magnetic moment, we calculate 
a large intrinsic anomalous Hall effect (AHE) of about 300 S/cm. It derives 
from the Berry curvature distribution of the Weyl points, which are only 
14 meV away from the Fermi level and isolated from trivial bands. 
Different from antiferromagnets Mn$_3X$ ($X$= Ge, Sn, Ga, Ir, Rh, and Pt),
where the AHE originates from the non-collinear magnetic structure, the
AHE in Ti$_2$MnAl stems directly from the Weyl points and is topologically   
protected. The large anomalous Hall conductivity (AHC) together with a 
low charge carrier concentration should give rise to a 
large anomalous Hall angle. In contrast to the Co-based \textit{ferromagnetic} 
Heusler compounds, the Weyl nodes in Ti$_2$MnAl do not derive from nodal lines 
due to the lack of mirror symmetries in the inverse Heusler structure. Since 
the magnetic structure breaks spin-rotation symmetry, the Weyl nodes are stable 
without SOC. Moreover, because of the large separation between Weyl points of 
opposite topological charge, the Fermi arcs extent up to $75\%$ of the reciprocal 
lattice vectors in length. This makes Ti$_2$MnAl an excellent candidate for 
the comprehensive study of magnetic Weyl semimetals. It is the first example 
of a material with Weyl points, large anomalous Hall effect and angle despite 
a vanishing net magnetic moment. 
\end{abstract}

\maketitle

\textit{Introduction.}
Weyl semimetals (WSMs) are recently discovered topological
states in condensed matter physics, that have attracted
extensive interest due to their novel physics and
potential applications in spintronics~\cite{Wan2011,volovik2003universe,murakami2007,Balents2011,Yan2017,Armitage2017}. In Weyl semimetals,
conduction and valence bands linearly touch each other
in three-dimensional momentum space at the Weyl points.
The Weyl points behave as monopoles of the Berry curvature
field with positive and negative (or left and right handed)
chirality, which leads to non-closed surface surface Fermi arcs connecting
a pair of Weyl points with opposite chirality~\cite{Xu2015TaAs,Lv2015TaAs,Yang2015TaAs,Liu2016NbPTaP,Xu2015NbAs,Belopolski2016NbP,Xu2016TaP,Souma2015NbP,Inoue2016, Batabyal2016, Zheng2016,Soluyanov2015WTe2,Sun2015MoTe2,Klaus_2016_TaIrTe4, Haubold_2017}.
Besides topological surface Fermi arcs, WSMs also exhibit exotic
transport phenomena, such as the chiral anomaly effect~\cite{Huang2015anomaly,Zhang2016ABJ,Wang2015NbP,Niemann2017}, the gravitational
anomaly effect~\cite{Gooth2017}, strong intrinsic anomalous and spin Hall effects~\cite{Burkov:2011de,Xu2011,Sun2016},
large magnetoresistance~\cite{Shekhar2015,Ghimire2015NbAs,Huang2015anomaly,Zhang2016ABJ,Wang2015NbP,Luo2015,Moll2015}, 
and even special catalytic activity~\cite{Rajamathi2017}.

The existence of Weyl points requires breaking of spin degeneracy, either
by breaking inversion or time reversal symmetry (or both)~\cite{Neumann1929}.
Since the Berry curvature is odd under time reversal, Weyl points in 
the latter case can generate a strong anomalous Hall effect (AHE).
Allthough the total magnetic moment vanishes, time-reversal 
symmetry is broken in compensated ferrimagnets. 
In contrast to antiferromagnets, bands in compensated 
ferrimagnets are spin split, which allows the
existence of Weyl points even without spin-orbit coupling (SOC). 
Additionally, the Weyl points in compensated ferrimagnets lead to a 
strong AHE despite a zero net magnetic moment. 
Different from the non-collinear antiferromagnets Mn$_3X$ 
($X$= Ge, Sn, Ga, Ir, Rh, and Pt)~\cite{Chen_2014,Nayak2016,Nakatsuji2015,Zhang_2017}, 
the AHE in a compensated magnetic WSMs is 
topologically protected. If the Weyl nodes are close to the 
Fermi level and are isolated from trivial bands, the charge carrier density 
is expected to be very low and a large anomalous Hall angle is naturally expected in these materials.

In this paper, we have theoretically studied an
ideal time reversal symmetry breaking WSM in
the inverse Heusler compound Ti$_2$MnAl with compensated 
magnetic structure, where the Weyl points are only
14 meV away from the Fermi energy. Owing to the large 
Berry curvature around the Weyl points, Ti$_2$MnAl has
an intrinsic anomalous Hall conductivity (AHC) of around 
300 $S/cm$ and despite a zero
net magnetic moment.
In Ti$_2$MnAl the Weyl points do not derive from mirror symmetry 
protected nodal lines, due to the lack of mirror planes in the 
inverse Heusler structure. Rather, the Weyl points exist without 
taking into account SOC, in contrast to the recently predicted 
Weyl-points in ferromagnetic Co-based magnetic Heusler compounds, 
which posses nodal lines close to the Fermi level.~\cite{Wang_Heusler_2016,Chang2016,Kubler_2016}.
Since the nodal lines normally disperse in energy, 
the Co-based magnetic Heusler
WSMs have higher bulk charge carrier density, which makes 
the detection of Weyl points related phenomena difficult.  
Due to the lower symmetry, Ti$_2$MnAl has a small charge 
carrier density of only $2\times10^{19}/cm^{3}$,
which is even smaller than the charge carrier density in 
NbP\cite{Klotz2016}. Therefore the Weyl points in Ti$_2$MnAl 
should be much easier to observe experimentally and a large 
anomalous Hall angleAHA is expected. Heusler compounds are 
ideal materials for the study of the interplay between 
magnetism and topology ~\cite{Stanislav2010, Claudia2007} 
due to their excellent tunability. According to our calculations, 
Ti$_2$MnGa and Ti$_2$MnIn have a similar electronic structure 
with Weyl points. However, only Ti$_2$MnAl has been successfully
grown in thin films with a high Curie temperature above 650 K.~\cite{Feng2015,Fang_2014}
Considering all these factors, Ti$_2$MnAl provides an excellent
platform for the study of magnetic Weyl physics and suggests a new
direction to obtain a strong AHE without a net magnetic moment.

\textit{Methods.}
To investigate the electronic band structure we have performed
$ab-initio$ calculations based on density functional theory (DFT)
using the full-potential local-orbital code (FPLO)~\cite{Koepernik1999} with a
localized atomic basis and full potential treatment. The exchange
and correlation energy was considered at the generalized gradient
approximation (GGA) level~\cite{perdew1996} and GGA plus U with
on site U of Ti-3d and Mn-3d orbital varying from 1 to 5 eV~\cite{Dudarev1998}.
The tight binding model Hamiltonian
was constructed by projecting the Bloch states onto atomic orbital
like Wannier functions. By using the tight binding model Hamiltonian,
we have calculated the surface state and intrinsic AHC.

\textit{Results.}
Ti$_2$MnAl has an inverse Heusler lattice structure with space group $F\bar{4}3M$
(No.216) [Fig. 1(a)]~\cite{Feng2015, Skaftouros2013}, which, without the
inclusion of spin-orbit coupling (SOC), shows a typical half metallic electronic
band structure, as shown in Fig.1 (c, d). Because of the compensated magnetic sublattices formed by
Ti and Mn, the net magnetic moment in Ti$_2$MnAl vanishes.
The spin-down channel forms an insulating gap of 0.5 eV, and the spin-up channel is
semimetallic, which is in good agreement with previous study~\cite{Skaftouros2013}. 
The spin-up bands in Fig. 1(c) show an easily identifiable 
band anti-crossing along the $L-W$ direction, indicating a band
inversion around the Fermi energy. Though there is a general
gap for the spin-up channel along the high symmetry directions,
its density of states is not zero at the Fermi level. Therefore, the spin-up
bands should cut the Fermi level in some lower symmetry directions.

\begin{figure}[htb]
\centering
\includegraphics[width=0.5\textwidth]{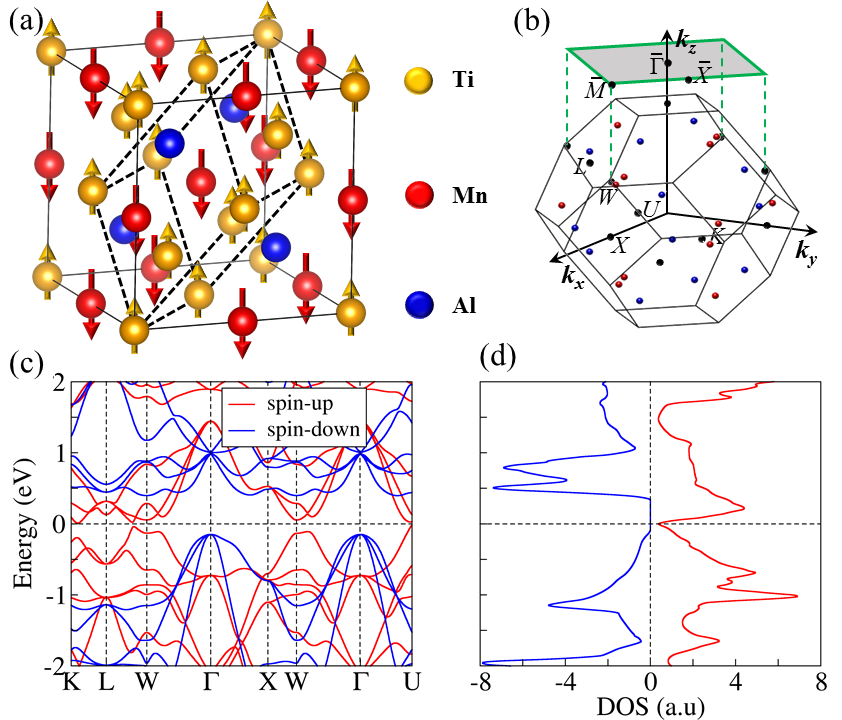}
   \caption{
(a) FCC lattice crystal structure of Ti$_2$MnAl. Spin polarizations of atom
Mn and Ti are along positive and negative $z$ direction, respectively. The primitive cell
is marked by the dashed line. (b) Three dimensional Brillouin zone
(BZ) and its two dimensional projection in (001) direction. Weyl
points with positive and negative chirality are presented by the
red and blue spheres. (c, d) Energy dispersion and density of
states without inclusion of SOC. The labels of high symmetry points
are given in (b).
}
\label{fig:lattice}
\end{figure}

Indeed, there is a linear crossing point away from the high-symmetry lines [Table \ref{tab:table1}].
Due to the six mirror planes $M_{\pm 110}$, $M_{1 \pm 10}$, and $M_{0\pm 11}$,
together with the three rotation axes  $C_{2,x}$, $C_{2,y}$,
and $C_{2,z}$, there are 12 pairs of linear crossing points in total.
This is completely different from the Co-based Heusler WSMs,
where the band structures possess nodal lines without considering
SOC. Calculating the Berry phases of these 12 pairs of linear crossing points,
we found that half of them have Chern number +1 and the other half Chern
number -1. Hence the chirality obeys the mirror symmetries, as
presented in Fig. 1(b). The energy dispersion of one pair of
Weyl points related by the mirror symmetry is given
in Fig. 2(a) and (c), from which we can see that the Weyl points
are just lying at the Fermi level.

\begin{table}[t]
  \caption{
Location of one pair of Weyl points (labled as $W_A$ and $W_B$). Without inclusion of SOC, they
are related to each other by the mirror plane , and all
the other Weyl points can be obtained via the symmetries of
$M_{\pm 110}$, $M_{1 \pm 10}$,  $M_{0\pm 11}$, $C_{2,x}$, $C_{2,y}$,
and $C_{2,z}$. SOC breaks the mirror symmetry as a perturbation. 
A relevant high symmetry point near the Weyl point is the $W$ point
(0, 0.5, 1.0).
The positions of Weyl points are described in the Cartesian coordinates in units of
$(\frac{2\pi}{a},\frac{2\pi}{a},\frac{2\pi}{a})$. 
}

  \label{tab:table1}
  \begin{tabular}{ccc}
  \hline
  \hline
           &         w/o SOC              &           SOC            \\
  $W_A$    &  (0, 0.4737, 0.7515)     &  (0, 0.4802, 0.7531)  \\
  $W_B$    &  (0.4737, 0, 0.7515)     &  (0.4680, 0, 0.7504)  \\ 
\hline
\hline
  \end{tabular}
\end{table}

\begin{figure}[htb]
\centering
\includegraphics[width=0.5\textwidth]{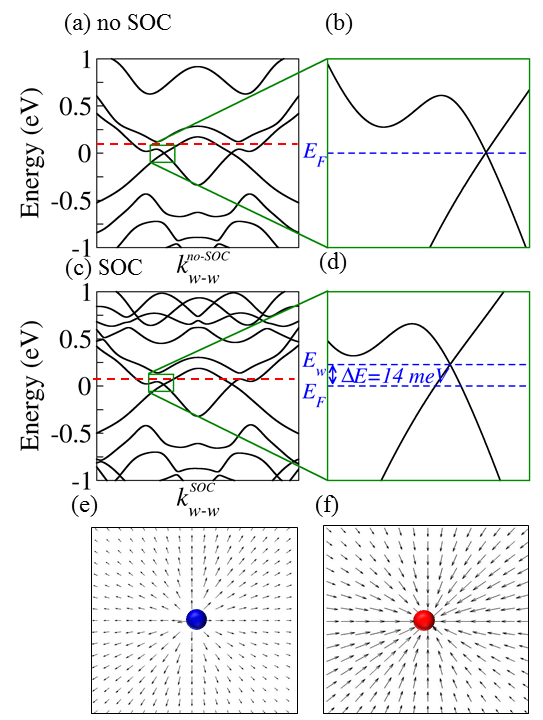}
   \caption{
(a-d) Energy dispersion of one pair of Weyl points for
(a, b) without and (c, d) with inclusion of SOC. The red
dashed lines in (a, c) are the reference lines indicating 
the breaking of the mirror symmetries if SOC is considered. 
(e, f) Berry curvature distribution around the Weyl points 
with positive and negative chirality.
}
\label{fig:wps}
\end{figure}

In order to completely understand the Weyl points in Ti$_2$MnAl, 
we need to take SOC into consideration, though SOC is not very 
strong in this compound, due to the light elements involved. 
In the presence of SOC, the symmetry of the system is reduced and the
details of band structures are dependent on the direction of the magnetization. 
To obtain the magnetic ground state, the total energies of 
magnetizations along the (001), (110), and (111) directions were compared. 
The net magnetic moments are still smaller than 1 $m\mu_{B}$ per formula unit
and the total energy difference between them is smaller than 0.1 meV, which is beyond 
the accuracy of DFT itself. Taking the magnetic polarization along (001) as an example, we analyzed the 
band structures further. Since SOC is not so strong in this compound, 
it just acts as a weak perturbation. 
As it is well known, the existence of Weyl points only need the lattice
translation symmetry, and they are robust as long as one pair of Weyl points with opposite 
chirality don't meet each other in the $k$-space.
The original position of one pair of Weyl points $W_A$ and $W_B$
and the energy dispersion slightly breaks the reflection symmetry $M_{\pm 110}$.
A comparison of the energy dispersions of one pair of Weyl points
before and after the inclusion of SOC is shown in Fig. 2(a-d). Without
the inclusion of SOC, the energy dispersion is mirror symmetric,
as shown by the red dashed line in Fig. 2(a). As long as SOC is taken into
consideration, the mirror symmetry is not preserved. The bands corresponding to $W_A$ 
cross the reference line (the red dashed line) two times, while the bands associated with 
$W_B$ cross it 4 times. Because of the perturbation from SOC,
the Weyl points are not at the charge neutrality point
any more, which shifts them above the Fermi level by around 14 meV,
see Fig. 2 (d). Though the SOC slightly shifts the Weyl points in
both momentum and energy, the chirality remains unchanged,
as shown in Fig. 2(e-f). 
We found that the existence of Weyl points is robust after inclusion of 
on site U for Ti-3d and Mn-3d orbitals from 1 to 5 eV.
 
A typical feature of WSMs is the topological surface Fermi arc state.
To calculate the (001) surface state, we considered an
open boundary condition with the half-infinite surface by using iterative
Green’s function method~\cite{Sancho1984,Sancho1985}. From the energy dispersion along the high-symmetry
line $\overline{X}-\overline{\Gamma}-\overline{M}$, surface bands connecting the bulk conduction and valence states
can be seen, 
which should be related to the Fermi arcs. Actually, the two projected
Weyl points on the (001) surface are very close to the
$\overline{\Gamma}-\overline{M}$ line (see Fig. 3 (c, d)), and therefore nearly linear
bulk band crossings can be observed along $\overline{\Gamma}-\overline{M}$.
Since Fermi arcs begin and end at the Weyl points with opposite
chiralities, we have chosen a special direction along the two projected
Weyl points (see Fig. 3 (c)). For the (001) surface, two Weyl points of the same 
chirality project onto each other. The corresponding energy dispersion, given
in Fig. 3(b), clearly shows see the linear dispersion of the
bulk Weyl points (labeled by the green and blue circles) with the Fermi arc
related surface bands terminating at the two Weyl points. 
Because of the dispersion of the Fermi arcs below the Fermi level, they could be detected by ARPES
and STM measurements.

By fixing the energy at the Weyl points, perfect Fermi arcs with tiny
bulk states can clearly be seen in Fig. 3(c). Dependent on the number of surface projected
Weyl points, the number of Fermi arcs terminated at each Weyl points differs.
Moreover, two long Fermi arcs extend around $75\%$ of the reciprocal
lattice vector, which is amongst the longest Fermi arc to be observed so far.
On shifting the energy down by 14 meV to the charge neutral point, there is slightly change in 
the surface state, and most of the Fermi arcs remain clearly detectable.
Therefore, the Weyl semimetal states in Ti$_2$MnAl lead to the 
existence of isolated surface Fermi arcs, and the long Fermi arc around the charge
neutral point should be easy to detect by surface techniques. 
Further, owing to the long Fermi arcs and small bulk charge carrier density, it is   
also easily to observe the Fermi arc induced quantum oscillation.

\begin{figure}[htb]
\centering
\includegraphics[width=0.5\textwidth]{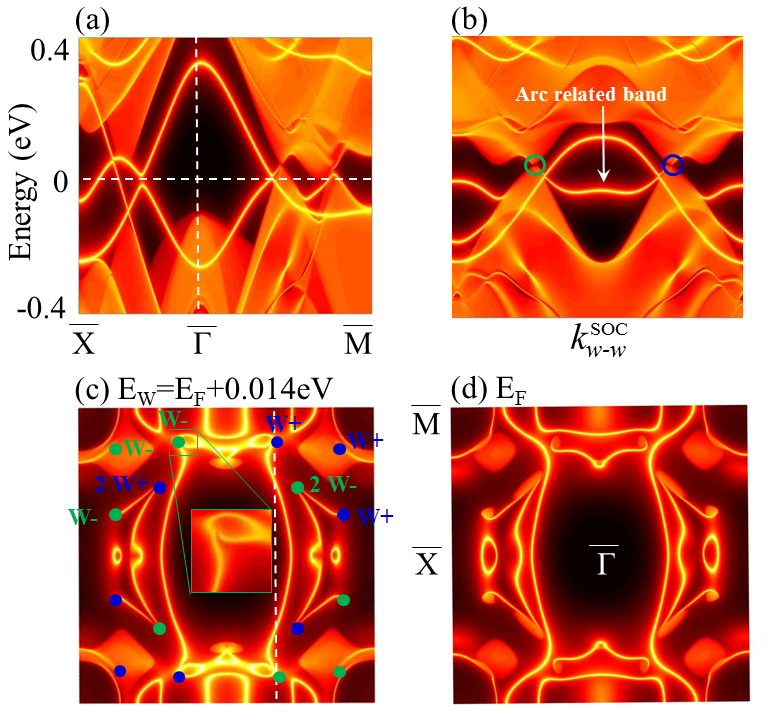}
   \caption{
Surface states of Ti$_2$MnAl terminated along the (001) direction.
(a) Surface energy dispersion along high symmetry lines of 
$\overline{X}-\overline{\Gamma}-\overline{M}$. (b) Surface 
energy dispersion crossing one pair of Weyl points.
The $k$-path is given in (c).
(c,d) Fermi surface with energy lying at Weyl points
and charge neutral point, respectively.
}
\label{fig:weyl}
\end{figure}

Owing to the large Berry curvature around the Weyl points, a strong AHE
is expected in magnetic Weyl semimetals. To investigate the AHE we have
computed the intrinsic AHC by the linear-response Kubo formula approach in
the clean limit~\cite{Xiao2010},
A $500 \times 500 \times 500$ $k$-grid in the BZ was used 
for the integral of the AHC.

\begin{figure}[htb]
\centering
\includegraphics[width=0.5\textwidth]{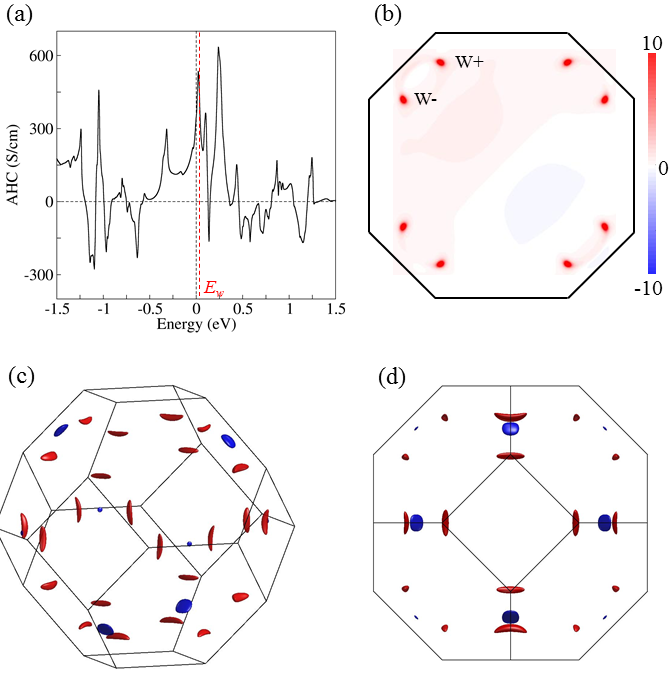}
   \caption{
(a) Energy dependent AHC. The peak value highlighted by red dashed line corresponds
to the Weyl-node energy. (b) Berry curvature
distribution in the $k_z$=0 plane. The eight hot spots are just
the positions of Weyl points. Color bars are arbitrary units.
(c, d) FSs distribution in the 3D BZ. Red
and blue FSs are hole and electron bubbles, respectively.
}
\label{fig:ahc}
\end{figure}

If the magnetization is along the $z$-axis, $\sigma_{xy}^{z}$ is the only
non-zero component of the AHC-tensor.
From our calculation, the intrinsic AHC is around 300
$S/cm$ at the charge neutrality point.
However, since the Weyl points are just above the Fermi level, the AHC should be sensitive to
the position of chemical potential and a large value is expected at the energy
of the Weyl points.
To investigate the effect of electron and hole doping, we calculated the energy
dependent AHC, see Fig. 4(a). The AHC increases sharply
when shifting the energy from $E_F$ to $E_w$,
and the AHC attains a peak value at $E=E_w$
with $\sigma_{xy}^{z}$=550 $S/cm$, implying that the large
AHC originated from the Weyl points, since the Berry curvature 
distribution of a Weyl point peaks sharply at the nodal energy. For further
confirmation of the effect of the Weyl points to the AHC, we
analyzed the Berry curvature distribution in $k$-space.
Fig. 4 (b) shows the Berry curvature distribution in the
$k_z$=0 plane with four pairs of Weyl points very close to it.
Except for the eight hot spots derived from the Weyl points,
there are barely other contributions to the AHC. The other
two high-symmetry planes $k_x$=0 and $k_y$=0 have almost the same
Berry curvature distribution. Therefore, the AHE in
Ti$_2$MnAl is mainly generated by the 12 pairs of Weyl
points, and hence it is topologically protected.

The AHA, defined as $\sigma_{xy}^{z}/\sigma_{xx}$, provides a
dimensionless measure of the strength of the AHE, which is the conversion
efficiency from longitudinal charge current to transverse
charge current. The longitudinal conductivity $\sigma_{xx}$ is proportional to the charge
carrier density and mobility. In comparison to metals,
the charge carrier density of semimetals differs by some orders of magnitude and therefore the
conductivity $\sigma_{xx}$ is lower as well.
Theoretically, the AHA can be estimated from the intrinsic AHC and charge carrier density. To
evaluate the charge carrier density we have analyzed the
FSs in three dimensions. As shown in Fig. 4(c,d), there
are 24 electron bubbles and 8 hole bubbles in the 1st BZ,
and all of them are very small. The charge carrier density
obtained from the volume of the FSs is around
$2\times10^{19}/cm^{3}$, a typical semimetallic value.
Though we cannot compute an accurate $\sigma_{xx}$, which
needs further experimental transport measurements,
a large AHA is expected from the small charge carrier density
and a relatively large AHC.
In the quantum anomalous Hall effect (QAHE), the AHA diverges, since the 2D bulk becomes insulating.
A magnetic material with a large bulk AHA could show a QAHE in thin
films due to quantum confinement~\cite{Xu2011}, and therefore the
study of thin film of Ti$_2$MnAl should be highly interesting.

\textit{Summary.}
In summary, we have theoretically predicted the magnetic
WSM and large AHC in the inverse Heusler compound
Ti$_2$MnAl with  Weyl points only 14 meV away from the
Fermi level. Because of the large Berry curvature from Weyl points and compensated
magnetic structure, the strong AHE exists even without a net magnetic moment.
Motivated by this findings, it would be intersting to investigate the topological
 properties of the other class of compensated ferrimagnets in the Heusler 
structure with 24 valence electrons ~\cite{Rolf2017}.
Compared to non-collinear antiferromagnets Mn$_3X$ ($X$=Ge, Sn, Ga, Ir, Rh, and Pt),
Ti$_2$MnAl has much lower charge carrier density while the AHC is of similar magnitude, from which a
large AHA and low energy consumption are also expected.
Due to the large separation of Weyl points with opposite
chirality, the  Fermi arc extend up to $75\%$ of the reciprocal lattice vectors in length.
This work not only provides an excellent magnetic Weyl semimetal for both transport
and surface study, but also suggests a new direction for obtaining a large AHE  without net magnetic moment.

\begin{acknowledgments}
This work was financially supported by the ERC Advanced Grant No. 
291472 `Idea Heusler', ERC Advanced Grant No. 742068--TOPMAT, 
EU FET Open RIA Grant No. 766566 grant (ASPIN), and 
Deutsche Forschungsgemeinschaft DFG under SFB 1143. 
Roberto Car and Lukas Muechler thanks to DOE grant DE-SC0017865.
LM would like to thank the MPI CPFS for its hospitality where part of the work was performed.
\end{acknowledgments}

\bibliography{TopMater}

\end{document}